\documentclass{ws-procs9x6}

\begin{document}

\title{Ferreting out the Fluffy Bunnies: Entanglement constrained by Generalized superselection rules} 

\author{\underline{Howard M. Wiseman}}

\address{Centre for Quantum Computer Technology, Centre for Quantum Dynamics, School of Science, Griffith University,  Brisbane,
Queensland 4111 Australia. \\
E-mail: H.Wiseman@griffith.edu.au}

\author{Stephen D. Bartlett} 
\address{Department of Physics, Macquarie University, Sydney, \\ New
  South Wales 2109, Australia.} 

\author{John A. Vaccaro}

\address{Division of Physics and Astronomy, University of
Hertfordshire, \\ Hatfield AL10 9AB, UK.}  


\maketitle

\newcommand{\beq}{\begin{equation}} 
\newcommand{\eeq}{\end{equation}}
\newcommand{\bqa}{\begin{eqnarray}} 
\newcommand{\eqa}{\end{eqnarray}}
\newcommand{\nn}{\nonumber} 
\newcommand{\nl}[1]{\nn \\ && {#1}\,}
\newcommand{\erf}[1]{Eq.~(\ref{#1})}
\newcommand{\dg}{^\dagger}
\newcommand{\rt}[1]{\sqrt{#1}\,}
\newcommand{\smallfrac}[2]{\mbox{$\frac{#1}{#2}$}}
\newcommand{\half}{\smallfrac{1}{2}} 
\newcommand{\bra}[1]{\langle{#1}|} 
\newcommand{\ket}[1]{|{#1}\rangle}
\newcommand{\ip}[2]{\langle{#1}|{#2}\rangle}
\newcommand{\sch}{Schr\"odinger } 
\newcommand{\hei}{Heisenberg } 
\newcommand{\bl}{{\bigl(}}
\newcommand{\br}{{\bigr)}} 
\newcommand{\ito}{It\^o }
\newcommand{\str}{Stratonovich } 
\newcommand{\dbd}[1]{\frac{\partial}{\partial {#1}}}
\newcommand{\du}{\partial}
\newcommand{\sq}[1]{\left[ {#1} \right]}
\newcommand{\cu}[1]{\left\{{#1} \right\}}
\newcommand{\ro}[1]{\left( {#1} \right)}
\newcommand{\an}[1]{\left\langle{#1}\right\rangle}
\newcommand{\st}[1]{\left| {#1} \right|}
\renewcommand{\implies}{\Longrightarrow}
\newcommand{\tr}[1]{{\rm Tr}\left[ {#1} \right]}

\newcommand{\cal}[1]{\mathcal{{#1}}}

\abstracts{
Entanglement is a resource central to quantum information (QI). In particular, entanglement shared between two distant parties allows them to do certain tasks that would otherwise be impossible. In this context, we study the effect on the available entanglement of physical restrictions on the local operations that can be performed by the two parties. We enforce these physical restrictions by generalized superselection rules (SSRs), which we define to be associated with a given group of physical transformations. Specifically the generalized SSR is that the local operations must be covariant with respect to that group. Then we operationally define the entanglement constrained by a SSR, and show that it may be far below that expected on the basis of a na\"ive (or ``fluffy bunny'') calculation. We consider two  examples. The first is a particle number SSR. Using this we show that for a two-mode BEC (with Alice owning  mode $A$ and Bob mode $B$),  the useful entanglement shared by Alice and Bob  is identically zero. The second, a SSR associated with the symmetric  group, is applicable to ensemble QI processing such as in liquid-NMR. We prove that even  for an ensemble comprising many pairs of qubits, with each pair described  by a pure Bell state, the entanglement per pair constrained by  
  this SSR goes to zero for a large ensemble. 
}

\section{Introduction}

Entanglement  is profoundly important in quantum information (QI) \cite{NieChu00}, and  has been much studied in recent years. 
Surprisingly, it is still a controversial topic, even for  pure states. 
For example, consider a Bose-Einstein condensate (BEC) 
containing $N$ atoms, suddenly split into two modes (say two internal states, or two wells). 

 S\o renson, Duan, Cirac and Zoller \cite{Sor01} 
claimed that  after some evolution (inter-mode cycling and intra-mode collisions), there would be entanglement 
 between the particles. They showed that this would be useful for precision measurement, exactly 
 as in non-condensed spin-squeezing \cite{Jul01}. Hines, McKenzie and Milburn \cite{HinMcKMil03}, however, criticized this characterization of entanglement, saying that ``the decomposition of the system \ldots 
into subsystems made up of individual bosons is not
physically realizable, due to the indistinguishibility of the
bosons within the condensates.'' 
 
The power of entanglement is seen most strikingly in its form as a {\em non-local resource}, allowing two distant parties to do certain tasks that would otherwise be impossible. In this context, one can understand the criticism by Hines {\em et al.} of the formalism of S\o renson {\em at al.} As an alternative, Hines {\em et al.} propose calculating the entanglement between the two modes, which is maximal immediately after the split. However, we argue, this entanglement  can only be accessed by the two parties if they violate fundamental conservation laws. Thus, in both of the above formalisms, the calculated entanglement is useless as a non-local resource for QI tasks.  {\em In this context},  both are examples of what Burnett has called {\em fluffy bunny entanglement} \cite{Bur03}. 
 
In this work we show how to calculate entanglement between two distant parties in the presence of physical restrictions on the local operations they can perform. The first step is to give an operational definition of entanglement in the presence of physical constraints \cite{WisVac03} (Sec.~II). The next is to define generalized superselection rules (SSRs) as a way to formulate a wide variety of constraints \cite{BarWis03} (Sec.~III). For such constraints an expression for the entanglement can then be derived and, for pure states, simply evaluated (Sec.~IV). The two main examples we have considered are the SSR associated with  particle number conservation \cite{WisVac03} (Sec.~V) and the SSR associated with  particle permutation covariance \cite{BarWis03} (Sec.~VI).  We conclude in Sec.~VII with a summary and a discussion of future investigations. 

\section{Operational Definition of Bipartite Entanglement} \label{sec:opdefent}

Say two distant parties, Alice and Bob, share some quantum system $S$ which for simplicity we will for the moment assume to be in a pure state $\ket{\psi_{AB}}$. Bipartite entanglement is a {resource} that enables Alice and Bob to do certain quantum tasks, such as teleportation \cite{NieChu00}, independent of the medium that holds that entanglement. We thus assume that, in addition to $S$, Alice and Bob each have a {conventional quantum register}, initially pure. 
We now define the entanglement in $S$ to be the
maximal amount of entanglement which Alice and Bob can produce between their quantum
registers by physically allowed local operations ${\cal O}$. 
With no physical constraints, the entanglement would simply be
\beq
 E(\ket{\psi_{AB}}) \equiv S(\rho_{A}),
 \eeq
where 
$S(\rho)=-\tr{\rho\log_{2}\rho}$, and $\rho_A = {\rm
Tr}_{B}[\ket{\psi_{AB}}\bra{\psi_{AB}}]/\ip{\psi_{AB}}{\psi_{AB}}$. Here Tr$_B$ denotes the trace over Bob's Hilbert space, and for convenience we are allowing for unnormalized states.
With physical constraints, it may be impossible to transfer this entanglement to the registers, so the constrained entanglement will be less than $E(\ket{\psi_{AB}})$. 
A simple example to which we will return is  the one-particle state $\ket{0_A}\ket{1_B} + \ket{1_A}\ket{0_B}$. This appears to contain $1$ ebit of entanglement. But to transfer this into the {conventional register} would require a {\tt SWAP} gate \cite{NieChu00}. Such a gate between this sytem and a conventional register is physically forbidden,  because it could act as follows:
\beq {\tt SWAP} \left( \alpha \ket{0_A^{\rm reg}} + \beta \ket{1_A^{\rm reg}} \right) \ket{0_A^{\rm sys}} 
=  \ket{0_A^{\rm reg}}  \left( \alpha \ket{0_A^{\rm sys}} + \beta \ket{1_A^{\rm sys}} \right).
\eeq 
Creating such a superposition, of the vacuum and one particle, would violate fundamental conservation laws, such as those for charge, lepton number {\em et cetera}, and hence is forbidden by superselection rules (SSRs).

\section{Physical Constraints as Generalized SSRs}

We define a  SSR to be a restriction (fundamental or practical) on the allowed {\em local} {operations} on a system. It is  {\em not}, in our view, a restriction on its allowed states. 
Note that ``operations" includes measurements as well as unitaries \cite{NieChu00}. 
More particularly, we define a {SSR} to be a rule associated with a {group} $G$ of {\em local} {physical transformations} $g$. The rule, which we denote the $G$-SSR,  is  that operations must be $G$-covariant. Here we define an operation ${\cal O}$ to be {$G$-covariant} if 
\beq
\forall \rho \textrm{ and } \forall g \in G\,, \;\mathcal{O}[\hat T(g) \rho \hat T^\dag(g)] = \hat T(g)[\mathcal{O}\rho]\hat T^\dag(g),
\eeq
where $\hat{T}(g)$ is a unitary representation of the group $G$. 

Traditionally \cite{Wic52} one talks of a SSR for an operator, rather than a SSR associated with a group of transformations.  For example, a SSR for {\em local} charge $\hat Q$  means that it is forbidden to create a superposition of states with different $Q$-values.
This can be derived from the conservation of global charge, an assumption that the initial state of the Universe had definite charge, plus the fact that global charge is the sum of local charges. However, as we will see, other SSRs may be practical rather than fundamental constraints. Note that it is pointless to talk of a SSR for global charge because that is a conserved quantity \cite{Wis03a}. Local charge is a {\em non-conserved} quantity so a SSR for it is meaningful.

The terminology ``SSR for $\hat Q$" or  ``$\hat{Q}$-SSR'' is compatible with our above definition if it is read as ``{SSR associated with the Lie group $G$ generated by $\hat Q$}." That is, the operations that cannot create local charge superpositions satisfy $\forall \rho \textrm{ and } \forall \xi \in \mathbb{R}, \mathcal{O}[e^{-i\xi \hat Q} \rho e^{i\xi \hat Q}] = e^{-i\xi \hat Q}[\mathcal{O}\rho]e^{i\xi \hat Q}$. 

If a SSR associated with $G$ is in force, then all experimental predictions are unchanged if a state $\rho$ is replaced by the state $\hat T(g)\rho \hat T\dg(g)$ for any $g \in G$. 
The {most mixed} state (that is,  the state containing no irrelevant information) with which $\rho$ is {physically equivalent} is  
\beq
{\mathcal{G}\rho \equiv \begin{cases} (\text{dim}\ G)^{-1}
  \sum_{g \in G}
  \hat T(g) \rho \hat T^\dag(g) \, ,  &\text{finite groups} \\ \int_{{}_{G}}
  \text{d}\mu_{{}_{\rm Haar}}(g)\, \hat T(g) \rho \hat T^\dag(g) \, ,  &\text{Lie
  groups}\end{cases}  .}
\eeq 
We call this the {$G$-invariant state}, as $\forall g \in G \,,
  \hat T(g) [\mathcal{G}\rho] \hat T^\dag(g) = \mathcal{G}\rho$. 


\section{Entanglement Constrained by a $G$-SSR}

Having precisely defined SSRs, we can now generalize (and specialize) our operational definition of entanglement to   
\beq \label{lab}
E_{G\textrm{-SSR}}(\rho_{AB}^{\rm sys}) = \max_{\mathcal{O}}\ E_{\rm D}\ro{{\rm
    Tr}_{\rm sys}\sq{
  \mathcal{O} \ro{\rho_{AB}^{\rm sys} \otimes \varrho_{AB}^{\rm reg} }}}.  
\eeq 
The generalization is that we have allowed for a mixed system state $\rho_{AB}^{\rm sys}$.
As a consequence the entanglement is not uniquely defined, so we have to specify the entanglement measure. Since we are interested in how much useful entanglement ends up in the registers, the entanglement of distillation $E_{\rm D}$ is a natural choice \cite{HorHorHor00}. 
The initial register state $\varrho^{\rm reg}_{AB}$ is still pure and separable.
The specialization of our previous definition is that the physical restrictions are those enforced by  a generalized SSR. That is, the operations ${\cal O}$ are {$G$-covariant} {\em local} operations. We can now prove the following \cite{BarWis03}: 

{\bf Theorem}: The SSR can be enforced by {\em removing all irrelevant information} from $\rho$ by the decoherence process $\rho \to {\cal G} \rho$. That is, 
\beq 
{E_{G\textrm{-SSR}}(\rho_{AB}) = E_{\rm D} \ro{{\cal G}\rho_{AB}}.}
\eeq


\section{Example: Particle Number SSR} 

As for charge,  {\em global} conservation laws  and suitable initial conditions lead 
to a SSR for {\em local} particle number $\hat{N}$, which is however {\em not} conserved.  
Because they are not subject to a number SSR, quanta of excitation such as photons or excitons  {\em are  not} particles by our meaning of the word. But electrons, protons, and Rubidium atoms in a specified electronic state {\em are}. 
Note that the global conservation law need not be for particle number; the total number of Rb atoms in the universe is not conserved. However there are conservation laws for lepton number, baryon number and so on that ensure that there is a Rb atom number SSR.

In this case the the entanglement constrained by the $\hat N$-SSR is
\beq
E_{\hat{N}-{\rm SSR}}(\rho_{AB}) = E_{\rm D} ({\cal N}\rho_{AB})
 \eeq
where the operation ${\cal N}$ destroys coherence between
 eigenspaces of $\hat\Pi_n$, having different {\em local} particle number $n$:
\beq
{{\cal N}\rho_{AB} = \sum_n \hat\Pi_n \rho_{AB} {\hat\Pi}_n.}
\eeq
Here, ``local'' could mean either Alice's or Bob's; it makes no difference. 

Consider a simple pure-state example.  
We use the notation of separating the occupation numbers of Alice's mode(s) from those of Bob's mode(s) by a comma, as in $\ket{n_A,n_B}$. For a particle in a  mode split between Alice and Bob, the state is  $\ket{\psi} = \ket{0,1}+\ket{1,0}$: a superposition, apparently with one e-bit of entanglement. But the equivalent invariant state is 
$
{\cal N} \ro{ \ket{\psi}\bra{\psi} } = \ket{0,1}\bra{0,1}+\ket{1,0}\bra{1,0},
$
an unentangled mixture. 

For general  pure states (which we here assume to be normalized), the entanglement constrained by the $\hat N$-SSR is
\beq
{E_{\hat N-{\rm SSR}}(\ket{\psi_{AB}}) =  
   \sum_n \bra{\psi_{AB}} \hat\Pi_n\ket{\psi_{AB}} S\bigl( {\rm Tr}_B \left[ \hat\Pi_n \ket{\psi_{AB}}\bra{\psi_{AB}} \hat\Pi_n \right] \bigr) .}
\eeq
Some pure state examples are given in Table~\ref{tab1}. 

\begin{table}[h!]
\tbl{Entanglement of various states according to the measure of Hines {\em et al.}, S\o renson {\em et al.}, and the present work. \vspace*{1pt}}
{\footnotesize
\begin{tabular}{|l|l|c|c|c|c|}
    \hline
    Description & State & ${E_{\rm Hines}}$ & $ {E_{\rm S\o r.}}$ & ${E_{\hat N-{\rm SSR}}}$  \\ 
    \hline\hline
   split particle &$\ket{0,1}+\ket{1,0}$ & {1} & {$-$} &  {0}   \\
    \hline
    Hines-entangled ``BEC''& $\ket{0,2}+\sqrt{2}\ket{1,1}+\ket{2,0}$ &  {3/2} & {0} & {0}    \\
    \hline
    S\o renson-entangled ``BEC''&$\ket{1,1}$ & {0} & {1} & {0} \\
     \hline
    either-entangled ``BEC'' & $\ket{0,2}+\ket{2,0}$ &  {1} & {1} & {0}\\
    \hline
    2 split particles & $(\ket{0,1}+\ket{1,0})^{\otimes 2}$ & {2} & {1} & {1/2}
    \\  \hline
    Bell pair&$\ket{01,10}+\ket{10,01}$ & {1} & {2} & {1} \\
    \hline
    ?&$\ket{11,00}+\ket{00,11}$ & {1} & {2} & {0} \\
    \hline
    $M$ split particles&$(\ket{0,1}+\ket{1,0})^{\otimes M}$ & {$M$} & {?} &  {$\sim M$} \\
    \hline
\end{tabular}\label{tab1} }
\end{table}

We now discuss some properties of $E_{\hat N-{\rm SSR}}$ illustrated by these examples. The first is {\em super-additivity} \cite{WisVac03}:
\beq
E_{\hat N-{\rm SSR}}(\ket{\psi}\otimes\ket{\phi}) \geq
E_{\hat N-{\rm SSR}}(\ket{\psi})+E_{\hat N-{\rm SSR}}(\ket{\phi}).
\eeq
All standard measures of entanglement are {\em sub-additive} \cite{HorHorHor00}. 
One could attribute this anomaly to the fact that for identical particles one $\ket{\psi}$
is not truly independent of another $\ket{\phi}$. 

The second property is 
{\em asymptotic recovery of standard entanglement} \cite{WisVac03}.   
For a large number $C$ of copies of a state $\ket{\psi}$,  
\beq E_{\hat N-{\rm SSR}}(\ket{\psi}^{\otimes C}) \sim
C E(\ket{\psi}) - \frac{1}{2}\log_{2}(V_{\psi}C) + O(1),
\eeq
where $V_\psi$ is the variance in Alice's particle number for  a single copy. 
Thus
\beq
\lim_{C\to\infty} E_{\hat N-{\rm SSR}}\ro{ \ket{\psi}^{\otimes C}} 
/ E\ro{\ket{\psi}^{\otimes C}} = 1 .
\eeq 

\section{Example: Ensemble QIP}

Ensemble QI processing means (i) 
there are $N \gg 1$ identical {copies} of a system (``a molecule'') containing $M$ qubits, and (ii) all operations are {\em collective} (i.e.~affect each molecule identically). 
For example, in  NMR QIP \cite{Cor97} 
each molecule contains $M$ atoms having a spin-$\frac12$ nucleus. The collective 
operations use (in general) spatially uniform RF magnetic pulses for unitaries and  fixed external antennae for measurements. 
In liquid-NMR the molecules can only be prepared in highly  mixed states. 
We show that {\em even if pure states could be produced}, the above restrictions imply that the  useful {entanglement {\em per molecule} goes to zero} as $N \to \infty$. 

The restriction on operations ${\cal O}$ can be formulated as the SSR
\beq
 \mathcal{O}[\hat T(p) \rho \hat T^\dag(p)] = \hat T(p)[\mathcal{O}\rho]\hat T^\dag(p)
 \eeq
Here {$p$ is a permutation} of the $N$ molecules and $\hat T(p)$ is the unitary operator that implements that permutation. We call this the $S_N$-SSR, as the $N!$ permutations $p$ form a group called the {\em Symmetric group} $S_N$.
We define the $S_N$-invariant (that is, randomly permuted) state 
\beq
\mathcal{P} \rho \equiv \frac{1}{N!} \sum_{p \in S_N} \hat T(p) \rho
  \hat T^\dag(p). 
\eeq

To understand the above, consider a simple example. 
Say $M=2$ (two nuclei, $A$ and $B$, per molecule) and $N=2$ (there are two molecules, 1 and 2), and the state is 
\beq
 \ket{\psi} =  \ket{\uparrow_A^1 \uparrow_B^1}\ket{\downarrow_A^2 \downarrow_B^2}.
 \eeq
We consider that the $A$s belong to Alice and the $B$s to Bob, and the $S_2$-SSR applies independently to Alice and to Bob. Now if Alice's local operations (acting only on nucleus $A$)  cannot distinguish molecules 1 and 2, then this state is {equivalent} to
\beq
{\hat T_A(1\leftrightarrow 2)\ket{\psi} = \ket{\downarrow_A^1 \uparrow_B^1}\ket{\uparrow_A^2 \downarrow_B^2}.}
\eeq
Under the action of ${\cal P}_A$ (or ${\cal P}_B$), $\ket{\psi}$  goes to an {equal mixture} of these two states, and all correlations are lost. 

Now consider a more  interesting example, 
where $N=2J$  and each molecule of the above sort  is prepared in a {\em pure} Bell state
\beq
{\ket{\Phi} = |\uparrow_A \uparrow_B\rangle  + |\downarrow_A \downarrow_B\rangle.}
\eeq
How much entanglement do Alice and Bob have at their disposal?   
 The na\"ive answer (no restrictions) is $N$ ebits  --- $1$ per molecule. By contrast, the constrained entanglement is  \cite{BarWis03}
\beq E_{S_N\textrm{-SSR}} \bigl( \ket{\Phi}^{\otimes N}  \bigr)  = 
 \sum_{j=0}^{J} \frac{(2j+1)^2}{2^{2J}(J+j+1)} 
\binom{2J}{J-j} \log_2 (2j+1) \sim \half \log_2 N \nn
\eeq
so the entanglement per molecule goes to zero as $N \to \infty$.   

\section{Discussion}

In this work we have argued as follows. Bipartite entanglement is a resource that enables the two parties to do certain quantum tasks, independent  of the medium that holds it. 
For many systems there are restrictions upon the physical operations, so na\"ive calculations of entanglement may over-estimate it. For such systems we  operationally define the entanglement as the amount of distillable entanglement that can be produced between two conventional (i.e.~unrestricted) registers. If the restrictions can be formulated as a generalized superselection rule (which we have defined) then we can derive an explicit  expression for this entanglement. 

We have considered two SSRs in detail, those associated with the group generated by local particle-number, and the group of  local permutations of particles. We have applied the first to a two-mode BEC (with Alice owning   mode $A$ and Bob mode $B$), and find that the entanglement shared by Alice and Bob  is identically zero.  We have applied the second to a {\em pure}  NMR ``ensemble Bell-state"  (with Alice owning nucleus $A$ and Bob nucleus $B$), and find that their entanglement  per molecule  is asymptotically zero. 

There are many other aspects of our work discussed in the papers ~\cite{WisVac03,BarWis03}. Our work also opens up many avenues of future investigation. First, there are the relations with reference frames and quantum communication \cite{Bar03}, and with quantum nonlocality \cite{TanWalCol91}. Second, there is the question of how to treat   physical restrictions not expressible as a SSR (as defined by us). 
In particular, QIP in NMR is {\em more restricted} than implied by our SSR because there are no controllable inter-molecular interactions. Thirdly, there is an apparent duality between our conclusion, that  in the presence of a SSR, a {\em nonseparable} $\rho$ {\em does not  imply} that the system is {\em entangled}, with the conclusion of Verstraete and Cirac \cite{VerCir03}, that 
  in the presence of a SSR, a {\em separable} $\rho$  {\em does not  imply} that the system is {\em locally preparable}. 
  
It is clear that
SSRs place severe contraints on QI processing, and our operational definitions
of SSRs and entanglement constrained by them provide a new understanding
and a valuable tool to QI science.
 
 \section*{Acknowledgments}
This work was supported by the Australian Research Council.


\begin{thebibliography}{0}

\bibitem {NieChu00} M. A. Nielsen and I. L. Chuang, \textit{Quantum
    Computation and Quantum Information,} (Cambridge University Press,
    Cambridge, 2000).

\bibitem{Sor01} A. S\o renson, L.-M. Duan, J. I. Cirac and P. Zoller, Nature {\bf 409}, 63 (2001).

\bibitem {Jul01} B. Julsgaard \textit{et al}, Nature 
  \textbf{413}, 400 (2001).

\bibitem{HinMcKMil03} A. P. Hines, R. H. McKenzie, and G. J. Milburn, Phys. Rev. A {\bf 67}, 013609 (2003).

\bibitem{Bur03} K. Burnett, personal communication.

\bibitem{WisVac03} H. M. Wiseman and J. A. Vaccaro,  Phys. Rev. Lett. {\bf 91}, 097902 (2003).
 
\bibitem{BarWis03} S. D. Bartlett and H. M. Wiseman, Phys. Rev. Lett. {\bf 91}, 097903 (2003).
 
 \bibitem {Wic52} G. C. Wick \textit{et al}, Phys. Rev. \textbf{88},
  101 (1952).
 
 \bibitem{Wis03a} H. M. Wiseman, Proceedings of SPIE {\bf 5111} 
  Fluctuations and Noise in Photonics and Quantum Optics,
  Eds. D. Abbott, J. H. Shapiro, and Y. Yamamoto (SPIE, Bellingham,
  WA, 2003), pp 78-91. \texttt{quant-ph/0303116}.
 
 \bibitem{HorHorHor00}
M. Horodecki, P. Horodecki, and R. Horodecki
Phys. Rev. Lett. {\bf 84} 2014 (2000).
 
 \bibitem {Cor97} D. G. Cory {\em et al.}, Proc. Natl. Acad. Sci.
    USA \textbf{94}, 1634 (1997); N. Gershenfeld and I. L. Chuang,
    Science \textbf{275}, 350 (1997).

\bibitem {Bar03} S. D. Bartlett, T. Rudolph, and R. W. Spekkens, 
  Phys. Rev. Lett. {\bf 91}, 027901 (2003).

\bibitem{TanWalCol91}
S. M. Tan, D. F. Walls, and M. J. Collett,
Phys. Rev. Lett. {\bf 66}, 252 (1991).
  
\bibitem {VerCir03} F. Verstraete and J. I. Cirac,
Phys. Rev. Lett. {\bf 91}, 010404 (2003)
 
 
\end{thebibliography}
\end{document}